\documentclass[sn-mathphys-num]{sn-jnl}
\usepackage{graphicx}%
\usepackage{multirow}%
\usepackage{amsmath,amssymb,amsfonts,colordvi,mathrsfs,bm,verbatim,dcolumn,epsfig,subfigure,float}%
\usepackage{amsthm}%
\usepackage{mathrsfs}%
\usepackage[title]{appendix}%
\usepackage{xcolor}%
\usepackage{textcomp}%
\usepackage{manyfoot}%
\usepackage{booktabs}%
\usepackage{algorithm}%
\usepackage{algorithmicx}%
\usepackage{algpseudocode}%
\usepackage{listings}%
\usepackage{romannum}
\usepackage[version=4]{mhchem}
\usepackage{mathtools}
\usepackage{times}
\usepackage{siunitx}

\theoremstyle{thmstyleone}%
\theoremstyle{thmstyletwo}%

\theoremstyle{thmstylethree}%

\begin{document}

\title[Article Title]{Frequency-Domain Denoising-Based \textit{in Vivo} Fluorescence Imaging}

\author[1]{\fnm{XuHao} \sur{Yu}}\equalcont{These authors contributed equally to this work.}
\author[2,3]{\fnm{RongYuan} \sur{Zhang}}\equalcont{These authors contributed equally to this work.}
\author[4]{\fnm{Zhen} \sur{Tian}}
\author[2]{\fnm{Yixuan} \sur{Chen}}
\author[1,5]{\fnm{JiaChen} \sur{Zhang}}
\author[6]{\fnm{Yue} \sur{Yuan}}
\author*[2]{\fnm{Zheng} \sur{Zhao}}
\author*[2]{\fnm{Ben Zhong} \sur{Tang}}
\author*[1,5]{\fnm{Dazhi} \sur{Hou}}

\email{dazhi@ustc.edu.cn}
\email{tangbenz@cuhk.edu.cn}
\email{zhaozheng@cuhk.edu.cn}

\affil[1]{International Center for Quantum Design of Functional Materials, Hefei National Research Center for Physical Sciences at the Microscale, University of Science and Technology of China, Anhui 230026, China}
\affil[2]{Guangdong Basic Research Center of Excellence for Aggregate Science, School of Science and Engineering, Shenzhen Institute of Aggregate Science and Technology, The Chinese University of Hong Kong, Shenzhen (CUHK-Shenzhen), Guangdong 518172, China}
\affil[3]{Department of Urology, Jining NO.1 People’s Hospital, Shandong 272000, China}
\affil[4]{Department of Urology, The First Affiliated Hospital of Soochow University, Jiangsu 215006, China}
\affil[5]{Department of Physics, University of Science and Technology of China, Anhui 230026, China}
\affil[6]{Department of Chemistry, University of Science and Technology of China, Anhui 230026, China}


\abstract{The second near-infrared window (NIR-II, 900–1,880 nm) has been pivotal in advancing \textit{in vivo} fluorescence imaging due to its superior penetration depth and contrast. Yet, its clinical utility remains limited by insufficient imaging temporal-spatial resolution and the absence of U.S. Food and Drug Administration (FDA)-approved NIR-II contrast agents. This work presents a frequency-domain denoising (FDD)-based \textit{in vivo} fluorescence imaging technique, which can improve signal-to-background ratio (SBR) and signal-to-noise ratio (SNR) by more than 2,500-fold and 300-fold, respectively. The great enhancement yields a doubled penetration depth and a 95\% reduction in contrast agent dosage or excitation light intensity for mouse vascular imaging. Additionally, we achieved a SBR far exceeded the Rose criterion in the observation of tumor margins and vessels in mice using Indocyanine Green (ICG), demonstrating the feasibility of NIR-II surgical navigation with FDA-approved agents. Furthermore, a 600 Hz real-time video enables visualization of the entire contrast agent diffusion process within the mouse body and differentiation between arteries and veins. This innovative technique, characterized by exceptional sensitivity, efficiency, and robustness, presents a promising solution for clinical applications, particularly in NIR-II surgical navigation.}

\maketitle

\section*{Introduction}\label{sec1}

The second near-infrared window (NIR-II, 900–1,880 nm) fluorescence imaging, first achieved in 2009 \cite{DHJ-2009}, has now found widespread application in \textit{in vivo} imaging, including vascular imaging \cite{CXY-vascular-2018, Add-vascular-2023, WQB-vascular-tumor-organ-2014, LWY-Organ-review-2024}, tumor imaging \cite{WQB-vascular-tumor-organ-2014, PKY-tumor-2018, ZXB-tumor-2024, LWY-Organ-review-2024}, organ imaging \cite{WQB-vascular-tumor-organ-2014, LWY-Organ-review-2024} and even surgical navigation \cite{First-in-human-2020, Surgical-navigation-2022}. Benefiting from its longer wavelength, NIR-II exhibits weaker scattering and autofluorescence compared to the first near-infrared window (NIR-I, 700–900 nm), resulting in reduced background noise, enhanced penetration depth and resolution \cite{DHJ-review-2017}. In pursuit of optimal image contrast, the NIR-II has been further delineated into several sub-bands, with the long-wavelength range (\textgreater\ 1,300 nm) demonstrating exceptional imaging capabilities that have been extensively validated \cite{QJ-water-2021}.

At the \textit{in vivo} level, imaging resolution is heavily dependent on the imaging contrast agents and techniques utilized. To enhance the quality of imaging resolution, substantial efforts have been made to develop new imaging contrast agents and materials, as well as to improve imaging techniques \cite{WQB-review-2020}. A variety of NIR-II molecules and materials have been explored for their potential as NIR-II imaging contrast agents, with their performance being rigorously evaluated. These developed NIR-II imaging contrast agents include single-walled carbon nanotubes (SWCNTs) \cite{DHJ-carbon-review-2015}, quantum dots (QDs) \cite{WQB-vascular-tumor-organ-2014, WQB-Ag2Se-2013}, rare earth-doped nanoparticles (RENPs) \cite{Rear-earth-2013, Rear-earth-2016}, semiconducting polymer-based nanoparticles (SPNPs) \cite{DHJ-SPNPs-2014, CZ-SPNPs-2017}, small molecular dyes (SMDs) \cite{DHJ-SMDs-2016, ZF-SMDs-2018}, and luminogens with aggregation-induced emission characteristic (AIEgens) \cite{ZRY-DIPT-ICF-2024, Add-AIE-2024}. However, no NIR-II luminogens have yet received approval from the U.S. Food and Drug Administration (FDA), primarily due to concerns over potential toxicity and low photoluminescence quantum yield \cite{WQB-review-2020}, while the FDA-approved NIR-I luminogen, Indocyanine green (ICG) \cite{ICG-1960}, falls short in performance within the NIR-II \cite{CXY-ICG-bad-2020}. On the other hand, alternative fluorescence imaging techniques have made strides in enhancing imaging quality from various perspectives. For example, confocal \cite{DHJ-confocal-2018, DHJ-confocal-2017}, multiphoton \cite{TBZ-two-photon-2018, TBZ-three-photon-2020}, and light-sheet imaging \cite{DHJ-light-sheet-2019, ZXD-light-sheet-2023} significantly boost resolution by minimizing scattering effects through precise control of the focal point. Fluorescence resonance energy transfer (FRET) imaging \cite{ZF-FRET-2020, ZF-FRET-2020-2} increases the penetration depth by extending the emission wavelength, and fluorescence lifetime imaging (FLI) \cite{ZF-life-time-2018, ZF-life-time-2021} strengthens the detection of faint signals by analyzing the fluorescence decay times. However, these techniques face limitations such as limited scalability, prolonged acquisition times, complex operation, and high costs, which hinder their \textit{in vivo} and clinical applications. It's noteworthy that post-processing denoising techniques based on mathematical algorithms, particularly cross-correlation \cite{OLID-2008} and frequency domain filtering \cite{Frequency-2009}, have proven effective at capturing weak signals amidst strong backgrounds. These techniques showed efficiency in thermal imaging \cite{Saitoh-Lock-in-2018, HDZ-Mn3Sn-2024} and fluorescence microscopy and hold great potential for \textit{in vivo} imaging which have not been explored yet. 

In this work, we developed a frequency-domain denoising (FDD)-based \textit{in vivo} fluorescence imaging technique that significantly enhanced SBR and SNR, and was compatible with both photostable AIE nano contrast agents and rapidly metabolized ICG molecules. The enhancement of SBR and SNR by the FDD technique allowed us to achieve a doubling of the penetration depth in intralipid and a 95\% reduction in the required contrast agent dosage or excitation light intensity in mice with the AIE luminogen DIPT-ICF \cite{ZRY-DIPT-ICF-2024}. We employed the FDD technique in \textit{in vivo} mouse imaging using ICG in the ultra-long wavelength range of 1,400–1,700 nm, which achieved not only a seven-fold increase in the duration of distinguishable tumor imaging, but also the demarcation of vessels overlapping the liver. Notably, the SNRs far exceeded the threshold for clear imaging as defined by the Rose criterion \cite{Rose-1973}, essentially making ICG an applicable NIR-II contrast agent. Additionally, taking advantage of the exceptional weak signal extraction capability of the FDD technique, a 600 Hz video capturing the entire contrast agent diffusion process within the mouse body yielded the blood flow velocity and distinguished arteries and veins. We also achieved a threefold increase in frame rate without sacrificing SBR or SNR, which effectively suppressed motion blur caused by spatial shifts during prolonged exposure. This FDD-based fluorescence imaging system, notable for its sensitivity and efficiency, not only holds the potential to accelerate the approval process of NIR-II contrast agents by a two-order reduction in imaging toxicity, but also offers a promising solution for fluorescence imaging in clinical healthcare and surgical navigation by enhancing the performance of ICG in NIR-II.

\section*{Results}\label{sec2}
\textbf{The FDD-based \textit{in vivo} fluorescence imaging technique.} Improving imaging contrast fundamentally lies in the enhancement of SBR and SNR, which are defined as follows: 
\begin{equation}
SBR=\frac{Signal-Background}{Background}.
\end{equation}
\begin{equation}
SNR=\frac{Signal-Bckground}{\textit{\text{Standard deviation of Background}}}.
\end{equation}
The visibility of signal depends on SBR, while the clarity relies on SNR, thus, noise reduction is the main challenge in imaging contrast enhancement, the focus of this study. Figure 1a shows the schematic of the FDD-based fluorescence imaging system. Fluorescent contrast agents were injected into mice via the tail vein, and the excitation light was delivered with a periodic modulation of intensity, inducing the contrast agents to emit periodic fluorescence. Simultaneously, using a near-infrared camera, we obtained an N-period time-series of fluorescence images with periodic amplitude variations, as shown in Fig. 1b. The contrast in each frame is limited by noise, yet it can be markedly enhanced by separating the signal and noise in the frequency domain. Figure 1d shows the temporal and frequency variations of fluorescence amplitude at a pixel on the mouse's ear. In the time domain, the fluorescence signal exhibits a periodic feature distinct from the non-periodic noise of comparable amplitude. In contrast, in the frequency domain, obtained from the Fourier transform, the noise shows a random distribution in the broad spectrum, while the fluorescence signal is located at the modulation frequency and overwhelms the noise therein. We extracted the fluorescence signal at the modulation frequency and performed parallel operations on each pixel, effectively filtering the noise, which significantly enhanced both SBR and SNR, yielding the denoised image as in Fig. 1c.\\
\\
\noindent \textbf{Imaging contrast improvement of the NIR-II contrast agent.} Figure 2a shows the original images (left column) and corresponding amplitude profiles (right column) from penetration tests of intralipid, using three capillaries filled with AIE luminogen (AIEgen) DIPT-ICF. Despite the contrast agent’s exceptional photostability and high luminescence intensity, the fluorescence signal attenuates rapidly with increasing intralipid coverage depth, dropping below the noise threshold beyond 6 mm. Figure 2b shows the FDD results, demonstrating a doubling of the penetration depth. The FDD process suppresses the noise from 3,000 photons to single-photon level, as shown in Fig. 2c and 2d (where the left axis represents pixel values and the right axis corresponds to photon counts), enabling the capture of signals as low as 3 photons (Fig. 2b, depth = 12 mm). Figure 2e and 2f show the SBR and SNR of the original and FDD images, respectively. Applying the FDD technique led to enhancements in both SBR and SNR across varying depths of intralipid coverage, with 734-fold and 98-fold improvements, respectively (Fig. 2e and 2f, right axes), achieved at the 6 mm penetration limit of the original image.

We further utilized DIPT-ICF for mouse whole-body vascular imaging in the 1,400–1,700 nm detection range \cite{QJ-water-2021, Water-2016, Water-2018} as shown in Fig. 3a-3d. In traditional spatially stable fluorescence imaging, the clarity is entirely determined by three degrees of freedom—exposure time, contrast agent dosage, and excitation light intensity, which must be reasonably allocated. Here, we set the exposure time to the upper detection limit of the camera and define the experimental resource as the product of the contrast agent dosage and excitation light intensity. Figure 3a and 3b show the original images under the same experimental resource, where the contrast agent dosage and excitation light intensity are 1/20 of each other, respectively. In the original images, only superficial large blood vessels beneath the skin of the mouse are visible. In contrast, the noise level is reduced by 98\% in the corresponding FDD images, as quantified in Fig. 3c and 3d, achieving a clear visualization of the mouse's entire vascular structure, particularly in the ear region. Leveraging the noise reduction capability of the FDD technique, both SBR and SNR are improved by more than an order of magnitude. The improvements are smaller than those shown in Fig. 2, due to fewer acquisition images. The results demonstrate the feasibility of high-contrast imaging under a 95\% reduction in contrast agent dosage and excitation light intensity, which can significantly reduce toxicity and cost in future clinical applications.\\
\\
\noindent \textbf{Imaging contrast improvement of ICG.} We performed tumor imaging in mice using ICG, the only FDA-approved contrast agent with an emission peak in NIR-I and a slight tail extending beyond 1,400 nm (Fig. S1) \cite{ICG-NIR-II-2018}. The ultra-long wavelength range of 1,400–1,700 nm was selected, yet far from the emission peak of ICG, resulting in decreasing imaging contrast, as shown in Fig. 4a, where the original images were captured at different time points. Ten minutes post-injection, vascular signals disappeared, while the tumor signal peaked with the margins distinguishable. Over time, the tumor signal weakened, with the margins becoming blurred by 30 minutes and the signal vanishing after 2 hours. In comparison, the corresponding FDD images presented in Fig. 4b not only demonstrate clearer tumor margins, but also reveal vascular structures on the mouse body, even those within the tumor. Additionally, the detectable durations of both tumor margins and tumor signals are extended to 2 hours and 8 hours, respectively, four times those of the original images. The prolonged effective time window for ICG can help to avoid repeated injections during surgical navigation, such as in adrenal tumor excision \cite{Adrenal-tumor-2016} and colorectal resection \cite{Colorectal-resection-2014}. Through the FDD technique, both tumor-to-background ratio (TBR) and tumor-to-noise ratio (TNR) are enhanced by at least 7-fold and 3-fold, respectively, as shown in Fig. 4c and 4d. The tumor was further excised, and its visible light image is shown in Fig. 4e. The original image of the excised tumor, displayed in Fig. 4f, is obscured by noise, making it challenging to discern the tumor's shape. In contrast, the FDD image presented in Fig. 4g demonstrates that the entire excised tissue emits fluorescence, corroborating its tumorous nature and aligning with the hematoxylin and eosin (H\&E) staining results shown in Fig. 4h. The results highlight the enhanced performance of ICG in NIR-II by leveraging the advantages of the FDD technique, demonstrating its potential to shorten procedural time by reducing the need for repeated injections, and to improve accuracy in surgical navigation due to enhanced SBR and SNR.

The slight emission tail of ICG in the NIR-II region allows for short-exposure invasive and long-exposure noninvasive localized vascular imaging at wavelengths above 1300 nm \cite{ICG-NIR-II-2018}. However, the short imaging window, caused by the rapid hepatic clearance of ICG, limits its application in medical diagnostics \cite{ICG-fast-2023, ICG-fast-2023-2}. Here, we performed mouse whole-body vascular imaging using ICG under the same parameters as in Fig. 4. Figure 5a shows the original image captured at the peak signal, where only a few abdominal and leg vessels can be identified amid noise coverage. Yet, they disappear within 8 minutes, as shown in Fig. 5b. Moreover, the stronger liver signals resulting from the rapid accumulation of ICG obscure the identification of other structures overlapping with the liver, further restricting the broader application of ICG \cite{ICG-liver-2021}. By utilizing the temporal variations in signal dynamics caused by hepatic ICG clearance, we replaced external laser modulation in the FDD processing, which not only reduced noise but also facilitated the differentiation of vessels passing through the liver, as shown in Fig. 5c. Figure 5d shows temporal variations at four different locations, where noise remains constant while signals change continuously. Additionally, the liver signal intensifies while the vascular signal diminishes, leading to subtle changes in the overlapping regions. The different temporal variations result in distinct corresponding amplitudes in the frequency domain, as shown in Fig. 5e, with the signal amplitude at the liver-vascular overlapping region lower than that of the liver, facilitating the identification of vessels overlapping with the liver (see Fig. 5a-5c, local magnifications of the liver and corresponding amplitude profiles). The noise is suppressed by the FDD technique, allowing vessels in the ear region (see Fig. 5a-5c, upper-left insets) to become visible, with a greater than 100-fold increase in SBR and SNR, as shown in Fig. 5f and 5g.

Notably, the SNRs in both FDD-based tumor and vascular imaging are 45 and 65, respectively, far exceeding the threshold for clear imaging as defined by the Rose criterion (which is 4 for the SNR used in this study) \cite{Rose-1973}, and are even comparable to those achieved by NIR-II contrast agents \cite{200Hz-heart-2021}, effectively making ICG a practical NIR-II contrast agent.
\\
\noindent \textbf{Video contrast and frame rate improvement of temporally varying signals.} To further demonstrate the capability of the FDD technique in capturing rapidly changing signals, we injected DIPT-ICF into mice via the tail vein and acquired the entire diffusion process of the contrast agent using an ultra-high frame rate of 600 Hz. Due to the extremely short exposure time (1.67 ms), the original video revealed almost no discernible details, as shown in Video S1. Given that the signal evolves over time, we applied the FDD technique with ‘internal modulation’ similar to that in Fig. 5. In this case, the FDD technique here was executed after each new frame during real-time imaging, ensuring invariant frame rates. Video S2 shows the FDD video, where the sequence of contrast agent flow through the organs is clearly visible, with a 5-fold increase in the time resolution compared with the highest one (100 Hz) in previous studies \cite{200Hz-heart-2021}. Figure 6a and 6b show the last frame from the original and the FDD videos, respectively. In comparison, the FDD image shows improved contrast, particularly in the ear region.

Interestingly, not only the amplitude but also the so-called ‘phase’ information can be obtained through the FDD technique. Mathematically, all time-varying signals can be decomposed into a series of trigonometric functions, with the fundamental frequency components, which can be fully described by amplitude and phase. Similar to the calculation of amplitude, we focus solely on the phase data in the Fourier-transformed spectrum, yielding the FDD phase image, as shown in Fig. 6c. Due to the varying timing of the fluorescent contrast agent's passage through different regions, phase differences arise, which in turn allow for the differentiation of organs. More precisely, Fig. 6d shows a zoomed-in view of Fig. 6a, 6b, and 6c (upper row), along with their corresponding line profiles (lower row). While only a single vessel can be identified in the original and FDD amplitude images, the FDD phase image reveals two distinct vessels, an artery (red) and a vein (blue). This distinction arises from the fact that the contrast agent reaches the artery earlier than the vein, which is reflected in the spatial phase difference. Specifically, Fig. 6e illustrates the filling process of the contrast agent in both the artery and vein, with the corresponding original and FDD videos provided in Video S3 and Video S4, respectively. These phase results are similar to those obtained from principal component analysis (PCA) \cite{WQB-vascular-tumor-organ-2014, 200Hz-heart-2021, DHJ-PCA-2018}. Through the FDD technique, both SBR and SNR are increased by more than 30-fold, as shown in Fig. 6f and 6g. Leveraging the noise reduction capability of the FDD technique, we further acquired the temporal variation in amplitude, as shown in Fig. 6h. Moreover, by differentiating the data in Fig. 6h, we even identified the signal peaks corresponding to the mouse’s heartbeats, as shown in Fig. 6i.\\
\\
\noindent \textbf{Video contrast and frame rate improvement of spatially varying signals.} We progressively altered the spatial position of the mouse injected with DIPT-ICF and captured original videos at frame rates of 1 Hz (Video S5) and 4 Hz (Video S6), as well as the FDD video at 4 Hz (Video S7). Extended Data Figure 1a shows the representative original image at a 1 Hz frame rate, where, despite the strong signal, severe motion blur is observed due to the prolonged exposure time. By increasing the imaging frame rate to 4 Hz, the motion blur decreased, yet the signal diminished, as shown in Extended Data Fig. 1b. We further increased the sampling frequency to 16 Hz and set the laser modulation frequency to 4 Hz, enabling the execution of a single cycle of the FDD technique at 4 Hz, as shown in Extended Data Fig. 1c. As a result, both motion blur and noise were mitigated, which is expected to improve the smoothness of surgical navigation. Due to the presence of motion blur, both SBR and SNR became less effective. As a result, we turned to the Structural Similarity Index (SSIM), which combines image luminance, contrast, and structural information, to provide a more comprehensive assessment of image quality \cite{SSIM-2004}. The reference image utilized for SSIM calculation was an FDD image with same parameters and a constant spatial position. The calculated results demonstrate a 21-fold enhancement in SSIM with the FDD technique, as shown in Extended Data Fig. 1d.

\section*{Discussion}\label{sec3}
The FDD-based \textit{in vivo} fluorescence imaging improves both spatial and temporal resolution. This FDD technique can be applied to all types of fluorophores, which does not require any photoswitch features in the agents employed in previous works \cite{OLID-2008, Frequency-2009}. The three-fold extension of tumor margin duration achieved with ICG can eliminate the repeated injections and accelerate the lesion identification process, ultimately reducing the total operation time. Furthermore, though rapidly cleared by the liver, vascular signals from ICG were successfully separated from both the noise and the strong liver signals, highlighting the versatility and broad applicability of the FDD technique. Additionally, the three-fold increase in frame rate effectively suppressing motion blur, which holds significant potential to enhance the smoothness of surgical navigation promptly. So we expect the immediate clinical adoption of the FDD technique, particularly in ICG-based surgical navigation.

On the other hand, various contrast agents have been developed to obtain clear NIR-II images, yet none of them have received the approval mainly due to potential toxicity and low quantum yield \cite{WQB-review-2020}. The 95\% reduction in toxicity during the FDD NIR-II in vivo imaging offers the potential to expedite the approval process for the first NIR-II contrast agent. Moreover, the FDD technique facilitates a more than 10-fold increase in contrast by extending three-fold of total acquisition time, providing a viable solution for contrast agents with low quantum yield to achieve clear imaging, such as ICG tumor imaging in the region of 1,400–1,700nm. This also offers greater tolerance of quantum yield for the development of low-toxicity contrast agents.

\bibliography{sn-bibliography}

\section*{Methods}\label{sec5}
\noindent \textbf{Materials.}
All chemicals and reagents were purchased from commercial suppliers, and solvents used for chemical reactions were distilled prior to use. Air- and moisture-sensitive reactions were conducted in flame-dried glassware under a nitrogen atmosphere. Pluronic\textsuperscript{\textregistered} F-127 was obtained from Sigma-Aldrich, and ICG was purchased from TCI. Dulbecco’s Modified Eagle’s Medium (DMEM), fetal bovine serum (FBS), penicillin-streptomycin, phosphate-buffered saline (PBS), and trypsin were sourced from Thermo Fisher Scientific.

\noindent \textbf{Fabrication of the AIE nano contrast agent.}
The AIE fluorophore DIPT-ICF was synthesized as previously reported \cite{ZRY-DIPT-ICF-2024}. DIPT-ICF (1 mg) and Pluronic\textsuperscript{\textregistered} F-127 (3 mg) were each dissolved in tetrahydrofuran (THF) and combined to form a homogeneous solution, which was then slowly added dropwise into deionized (DI) water (9 mL) under gentle stirring. The resulting mixture was sonicated for 5 min in an ice bath using a Scientz-IID ultrasonic processor (Ningbo Scientz Biotechnology), followed by magnetic stirring for 12 h at room temperature. The solution was then dialyzed against DI water for 48 h. Finally, the obtained AIE nanoparticle suspension was concentrated to 1.0 mg mL$^{-1}$ and sterilized by filtration through a 0.22 \textmu m polyethersulfone (PES) membrane filter (Merck Millipore, Ireland) prior to use.

\noindent \textbf{Preparation and operation of mouse models.}
8-week-old female and male BALB/c nude mice (weighing 18–22 g) were purchased from the Guangdong Medical Laboratory Animal Center. All study protocols in this study were approved by the Institutional Animal Care and Use Committee (IACUC) of the Animal Experiment Center of the Chinese University of Hong Kong (Shenzhen, China).

Human renal cell carcinoma (RCC) 786-O cells were cultured under standard conditions and harvested during the logarithmic growth phase. Approximately 1 × 10$^{6}$ cells suspended in 100 \textmu L of serum-free medium were subcutaneously injected into the right gluteal region of BALB/c nude mice. Tumor growth was monitored every 2–3 days using a digital caliper, and tumor volume was calculated using the formula: 
\[
V = \frac{\text{length} \times \text{width}^2}{2}.
\] 
When the tumor volume reached approximately 1000 mm$^{3}$, mice were intravenously injected with ICG. Fluorescence imaging of the tumors was performed at various time points post-injection to monitor ICG accumulation and distribution. At the final time point, the subcutaneous tumors were surgically excised and subjected to ex vivo fluorescence imaging.

Contrast agents were injected into mice via the tail vein. Before imaging, mice were continuously administered inhaled isoflurane-O2 (3\% isoflurane with 97\% O2) for general anesthesia. All \textit{in vivo} experiments were performed using different mice across three separate trials to minimize incidental effects.

\noindent \textbf{Configuration of the imaging system and experiments.}
Excitation light at specific wavelengths (808 nm or 980 nm) was generated by a laser source (Connet Laser Technology). Two light homogenizers (Changchun New Industries Optoelectronics Technology) were used to achieve uniform and symmetric illumination of the target. The emitted fluorescence from the contrast agent was collected and focused by a lens (Navitar) onto the sensor of an InGaAs camera (FirstLight) to generate fluorescence images. Excitation light was blocked by long-pass filters (Thorlabs). A field-programmable gate array (FPGA) board (Anhui Luxet Infrared Technology), integrated with the FDD algorithm, was employed to modulate the laser output at a preset frequency and synchronize the camera acquisition. All captured frames were processed in real-time on the FPGA, and the resulting images were iteratively displayed on a monitor at the end of each acquisition cycle.

\noindent \textbf{Intralipid penetration measurement.}
Three capillaries (0.5 mm in diameter, spaced 1 cm apart) were filled with a 1 mg/mL solution of DIPT-ICF and embedded beneath a 1\% Intralipid phantom. The Intralipid thickness was varied from 0 to 12 mm in 1 mm increments. Excitation light at 980 nm with a power density of 32 mW/cm² was used. The camera exposure times were 25 ms, and the total FDD acquisition times were 80 s. A 1250-nm-LP filter was employed for emission signal collection.

\noindent \textbf{Mouse vascular imaging with DIPT-ICF.}
The excitation wavelength was 980 nm. 0.01 mg (0.05 mg/ml, 200 \textmu L) DIPT-ICF with 32 mW/cm² light power density and 0.2 mg (1 mg/ml, 200 \textmu L) DIPT-ICF with 1.6 mW/cm² light power density were used to acquire fluorescence images under low experimental resource conditions. A 1400-nm-LP filter was used for filtration. The camera exposure times were 10 s, and the total FDD acquisition times were 40 s. 

\noindent \textbf{Mouse tumor and vascular imaging with ICG.}
ICG (0.2 mg/mL, 200 \textmu L) was used to demonstrate the potential of the FDD technique in tumor and vascular imaging. The excitation wavelength was 808 nm, with a light power density of 32 mW/cm². The detection wavelength range was limited by a 1400-nm-LP filter. Tumor images were acquired at multiple time points post-injection: 10 min, 30 min, 1 h, 2 h, 4 h, and 8 h. Vascular imaging was performed by capturing the dynamic evolution of vascular signals over time. The camera exposure times were 4 s, and the total FDD acquisition times of tumor and vascular imaing were 320 s and 480 s, respectively.

\noindent \textbf{Real-time imaging of temporally varying signals.}
DIPT-ICF (1 mg/ml, 200 \textmu L) was used for real-time imaging. Excitation was provided by a 980 nm laser at a power density of 96 mW/cm², and a 1300-nm-LP filter was used to collect the emission signal. To capture the rapid signal fluctuations caused by contrast agent movement within the mouse, real-time imaging was conducted at a high frame rate of 600 Hz, the maximum supported by the camera, and synchronized with the injection of the contrast agent. After the acquisition of each new frame, an FDD calculation was performed based on all previously collected frames, maintaining a constant frame rate for the FDD-based real-time imaging. The total acquisition time was approximately 16 s.

\noindent \textbf{Real-time imaging of spatially varying signals.}
DIPT-ICF (1 mg/ml, 200 \textmu L) was used for real-time imaging. A 980 nm laser with a power density of 32 mW/cm² was used for excitation, and a 1400-nm-LP filter was employed for emission collection. Spatially varying signals were generated by manually moving the mouse. Real-time imaging was performed at frame rates of 1 Hz and 4 Hz, with total acquisition times of 4 s.

\noindent \textbf{Statistical Analysis.}
In this study, the original images refer to the averaged intensity of all frames acquired during the same period as the corresponding FDD acquisition, except for Fig. 5, Fig. 6 and Extended Data Fig. 1, where the signals were dynamic. The display range of the original images was adjusted to optimize visual clarity. Quantitative data are presented as mean values with error bars representing the range from minimum to maximum. Statistical analysis was performed using a two-sided Student's \textit{t} test. The levels of statistical significance were categorized as * \textit{P} \textless\ 0.05, ** \textit{P} \textless\ 0.01, and *** \textit{P} \textless\ 0.001.

\section*{Acknowledgments}\label{sec6}
The authors gratefully acknowledge financial support from the National Key Research and Development Program of China (2022YFA1403502 and 2023YFB3810001), the National Natural Science Foundation of China (12234017, 12074366 and 52333007), the Fundamental Research Funds for the Central Universities (WK9990000116), Shenzhen Key Laboratory of Functional Aggregate Materials (ZDSYS20211021111400001), and the Science Technology Innovation Commission of Shenzhen Municipality (KQTD20210811090142053 and JCYJ20220818103007014). The authors thank the supported by the USTC Center for Micro- and Nanoscale Research and Fabrication.

\section*{Author contributions}\label{sec7}
B.T., D.H. and Z.Z. conducted the project. X.Y., R.Z., D.H. and Z.Z. designed the experiments. X.Y., D.H. and J.Z. developed the imaging system. R.Z., C.Y. and Z.T. synthesized the fluorochromes. X.Y. and D.H. analysed the data and wrote the manuscript. Z.T. and Y.Y. helped the experiments.

\section*{Competing interests}\label{sec8}
The authors declare that they have no competing interests.

\section*{Additional information}\label{sec9}
\textbf{Supporting Information}\\
Supplementary Figure 1, Supplementary Table 1, Supplementary Videos 1-7\\
The frame rates of Videos S1-S4 were set to 30 Hz due to the large storage requirements of the original 600 Hz data.

\newpage
\begin{figure}[H]
\centering
\includegraphics[width=0.9\textwidth]{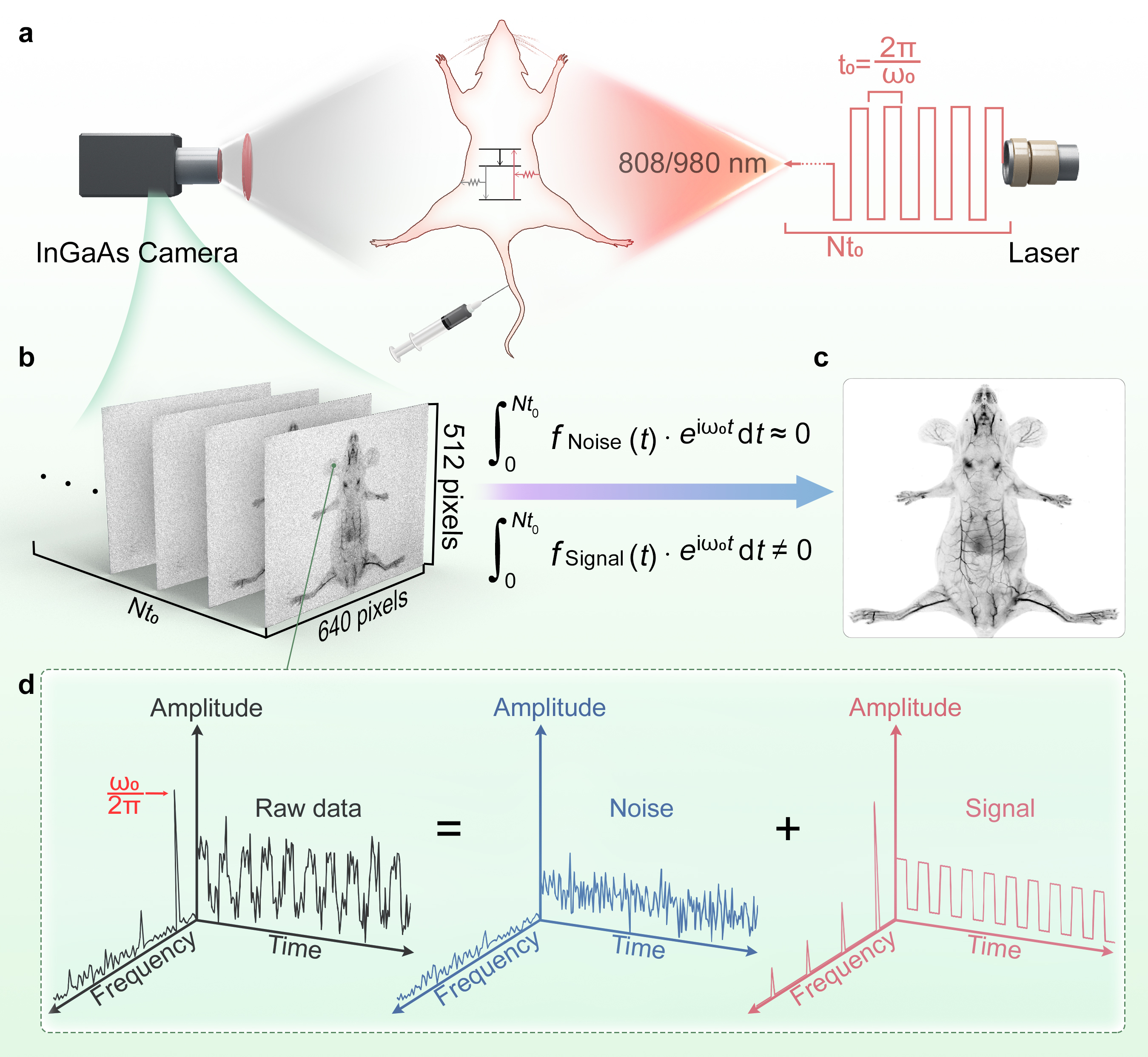}
\caption{\textbf{Principle of the FDD-based \textit{in vivo} fluorescence imaging measurements. a}, Setup of the FDD-based \textit{in vivo} fluorescence imaging system. \( t_0 \): Time of a sampling cycle. \(\omega_0\): Angular frequency of the FDD technique. N: Number of sampling periods. \textbf{b}, Images captured for the FDD calculations. \textbf{c}, Image calculated by the FDD technique. The expressions between \textbf{b} and \textbf{c} yield the amplitude projections of the signal and noise components at the fundamental frequency \(\omega_0\), respectively, where the signal is preserved and the noise is effectively attenuated to near zero. \textbf{d}, Fluorescence amplitude variations of a pixel on the mouse ear in the time and frequency domain. The results in the frequency domain are obtained by applying the Fourier transform to the time domain data.}\label{fig1}
\end{figure}

\newpage
\begin{figure}[H]
\centering
\includegraphics[width=0.9\textwidth]{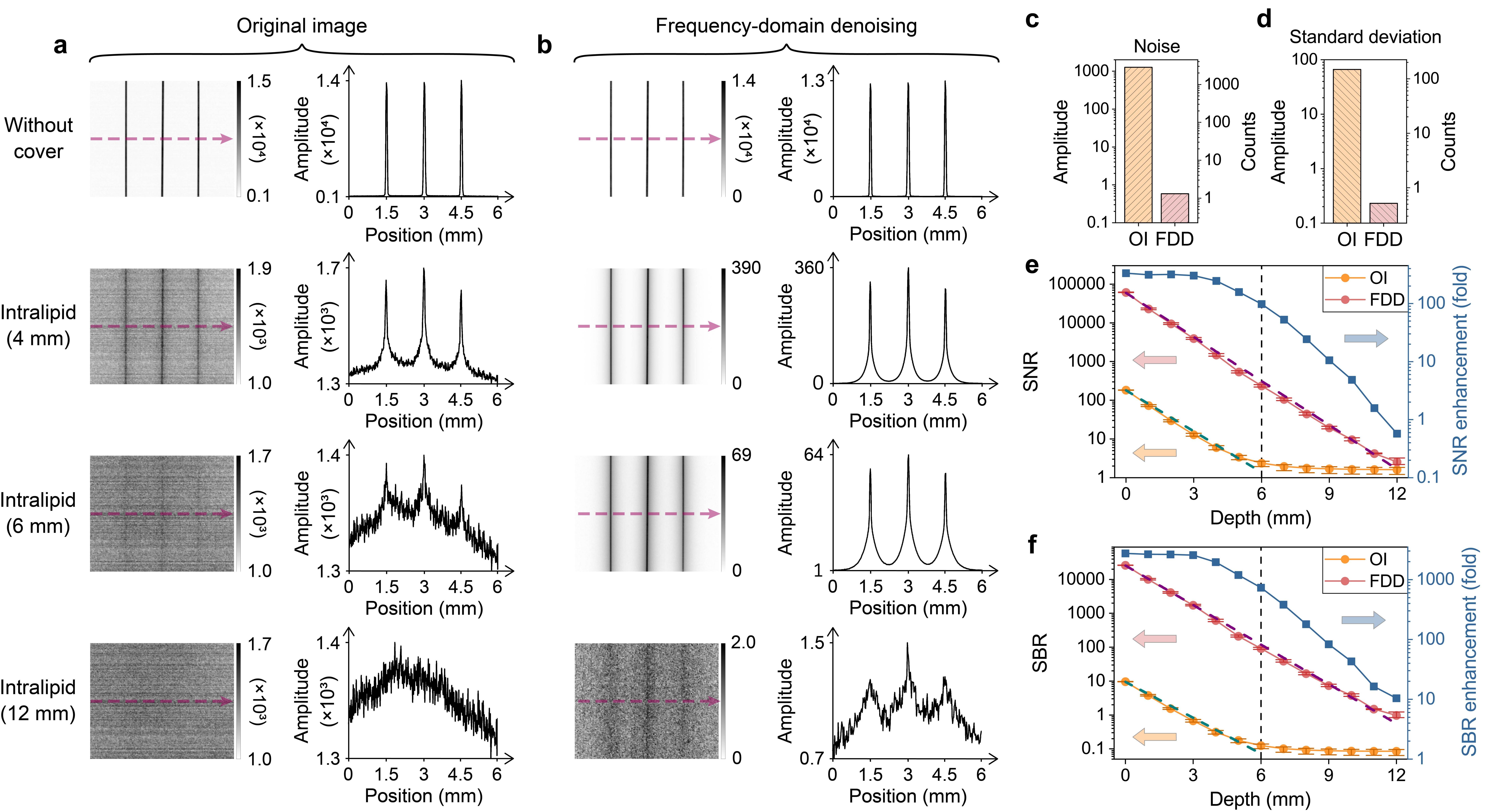}
\caption{\textbf{Measurements of Intralipid penetration depth by DIPT-ICF. a}, The original images under different depth intralipid overlay (left column) and corresponding amplitude profiles (right column). \textbf{b}, The corresponding FDD images (left column) and amplitude profiles (right column). The amplitude profiles are obtained by averaging all the rows of the corresponding images. \textbf{c, d}, Noise and standard deviation comparisons for the original and FDD images. The left y-axis represents pixel values, while the right y-axis shows the corresponding photon counts. Each unit of amplitude in the images captured by the camera corresponds to an average of 2.24 photons. \textbf{e}, SBR (left y-axis) and SBR enhancement (right y-axis) for the original and FDD images. \textbf{f}, SNR (left y-axis) and SNR enhancement (right y-axis) for the original and FDD images. The signals are the average of center columns of the three tubes. The backgrounds are the average of regions of interest (ROIs) selected from a relatively flat area distant from the signal. The statistical analyses are shown in Table S1.}\label{fig2}
\end{figure}

\newpage
\begin{figure}[H]
\centering
\includegraphics[width=0.9\textwidth]{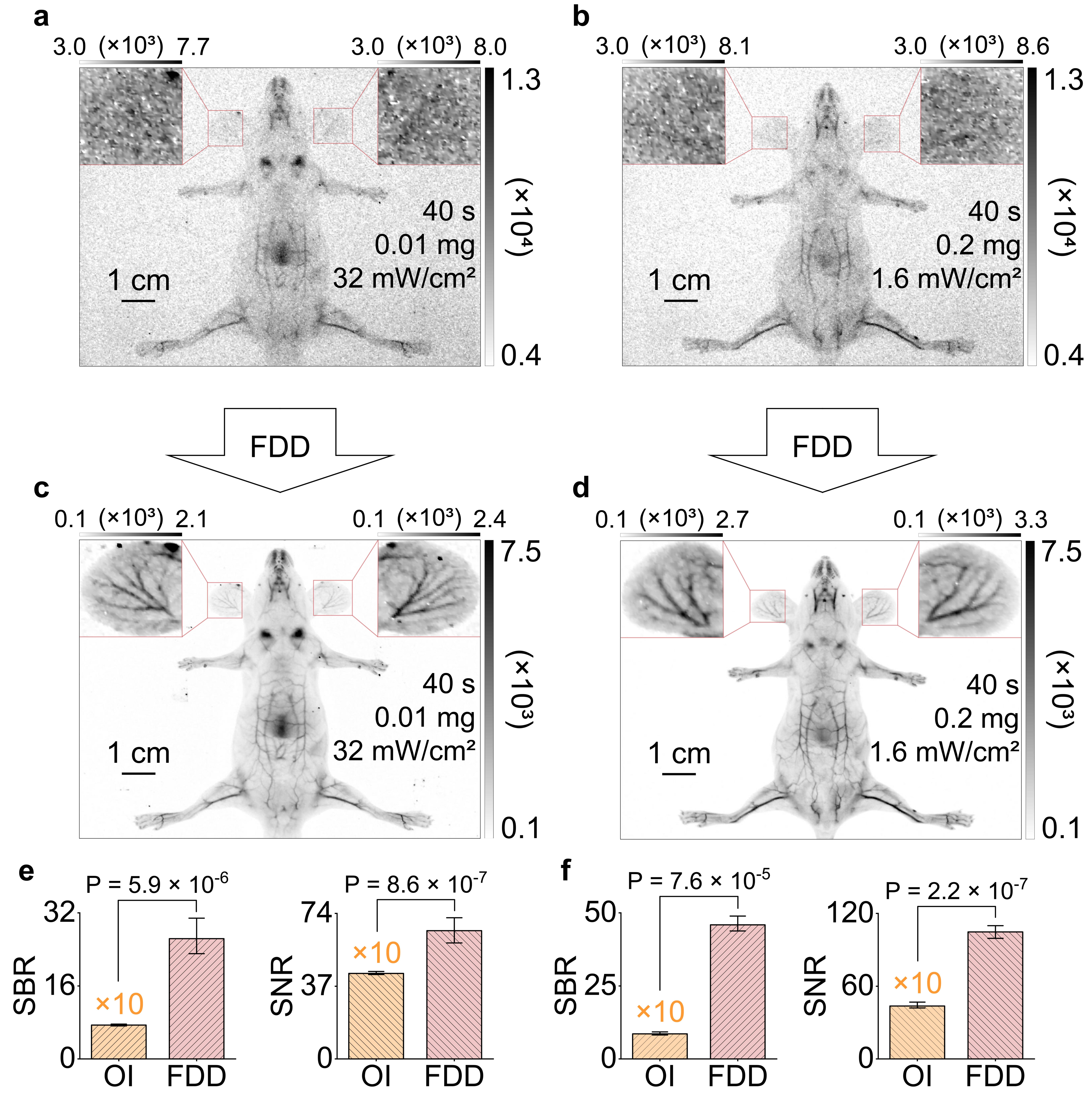}
\caption{\textbf{Florescence imaging of the mouse whole-body vascular by DIPT-ICF. a}, The original image. Dye dose = 0.01 mg. Excitation light power density = 32 mW/\(\text{cm}^2\). \textbf{b}, The original image. Dye dose = 0.2 mg. Excitation light power density = 1.6 mW/\(\text{cm}^2\). \textbf{c}, The FDD image corresponding to panel \textbf{a}. \textbf{d}, The FDD image corresponding to panel \textbf{b}. \textbf{e, f}, SBR and SNR of the original and FDD images. The signals are the averages of the maximum datas from the two ears of each mouse. The backgrounds are the averages of ROIs selected from relatively flat areas outside the mouse bodies.}\label{fig3}
\end{figure}

\newpage
\begin{figure}[H]
\centering
\includegraphics[width=0.9\textwidth]{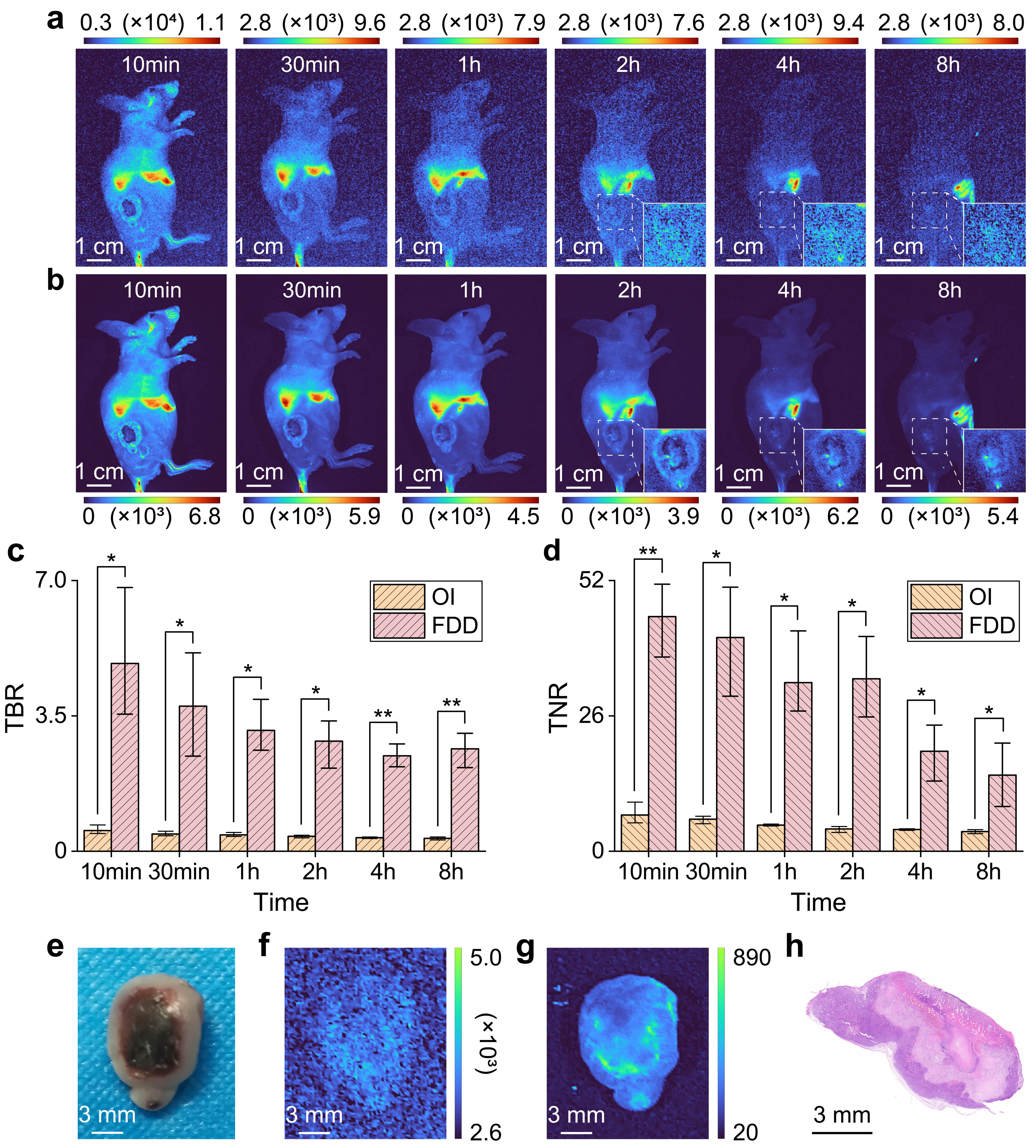}
\caption{\textbf{Florescence imaging of the mouse tumor by ICG. a, b}, Temporal changes in the original and FDD images. \textbf{c, d}, SBR and SNR of the original and FDD images. The signals are the averages of the maximum data from the tumors. The backgrounds are the averages of ROIs selected from relatively flat areas on the mice skin close to the tumors. \textbf{e}, Visible light image of the excised tumor. \textbf{f}, The original image of the excised tumor. \textbf{g}, The FDD image of the excised tumor. \textbf{h}, H\&E staining of the excised tumor.}\label{fig4}
\end{figure}

\newpage
\begin{figure}[H]
\centering
\includegraphics[width=0.9\textwidth]{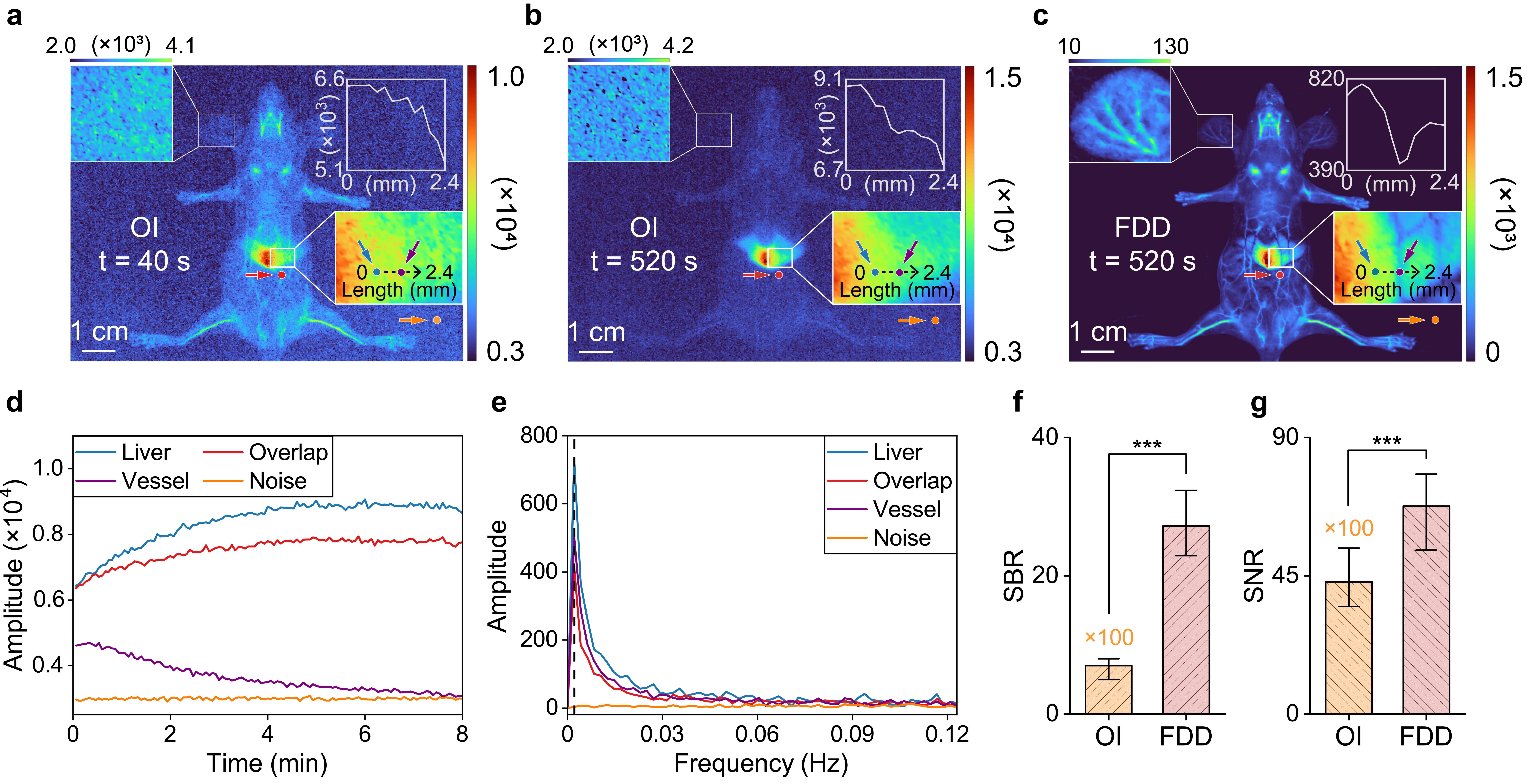}
\caption{\textbf{Florescence imaging of the mouse whole-body vascular by ICG. a}, The original image at the peak vessel signal. \textbf{b}, The original image captured after 8 minutes. \textbf{c}, The FDD image. Top left insets: magnification of the left ear. Bottom right insets: local magnifications of the live. Top right insets: amplitude profiles of the liver and overlapping vessel. \textbf{d}, Amplitude of a pixel in liver (green line), vessel overlapping liver (purple line), vessel (red line), and noise (yellow line) versus time. \textbf{e}, Fourier transform of panel \textbf{d}. The peak corresponds to the value of the corresponding point in panel \textbf{c}, and its frequency (dashed line) is the reciprocal of the acquisition time. \textbf{f, g}, SBR and SNR of the original and FDD images. The signals are the averages of the maximum data from the two ears of each mouse. The backgrounds are the averages of ROIs selected from relatively flat areas outside the mice bodies.}\label{fig5}
\end{figure}

\newpage
\begin{figure}[H]
\centering
\includegraphics[width=0.9\textwidth]{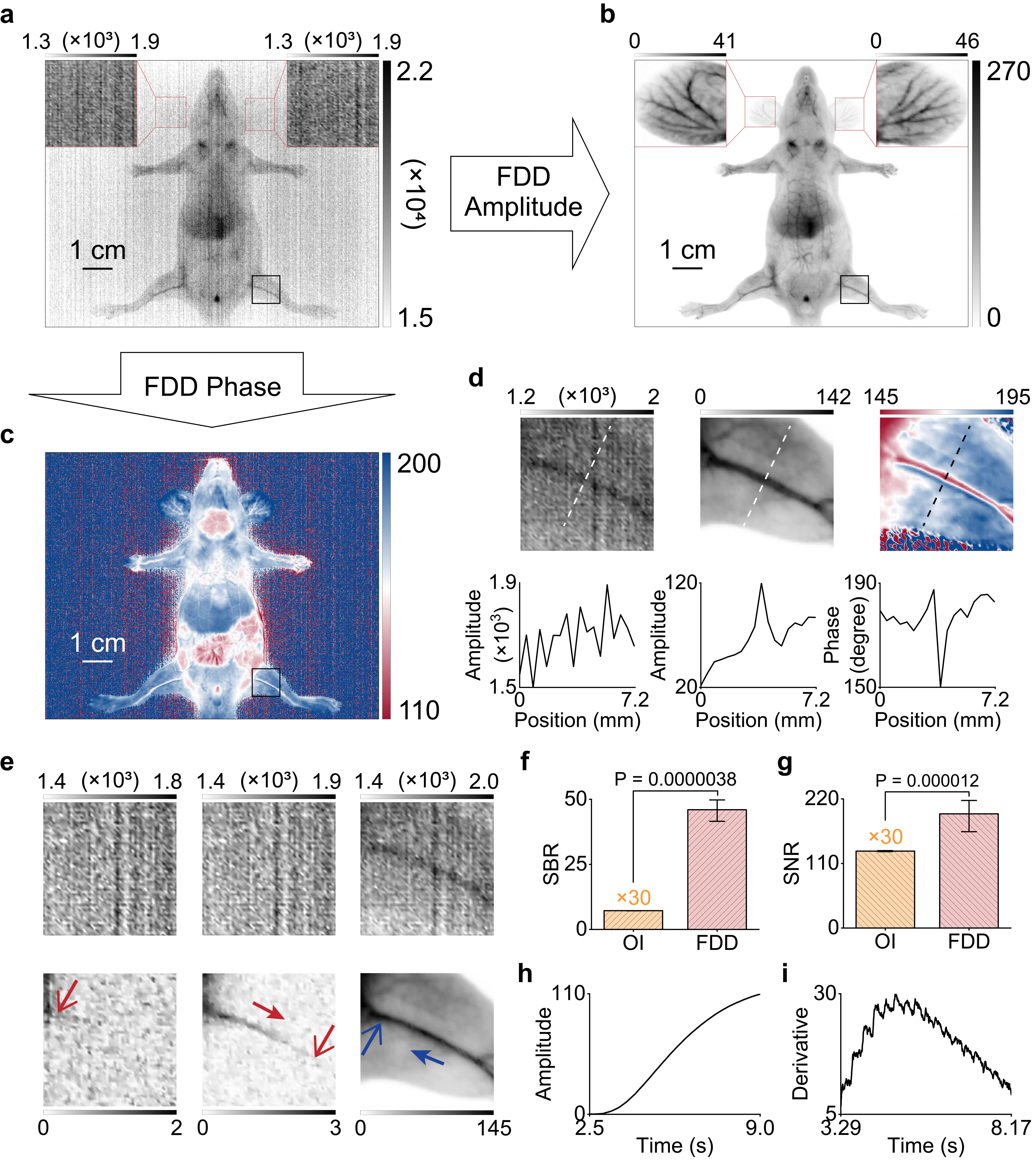}
\caption{\textbf{High spatio-temporal resolution whole-body vascular imaging. a}, The original image. \textbf{b}, The FDD amplitude image. \textbf{c}, The FDD phase image. \textbf{d}, Magnification of panels \textbf{a}, \textbf{b}, and \textbf{c} (upper row) and the corresponding intensity profiles (lower row). The amplitude images reveal only a single vessel, whereas the phase image distinguishes two vessels—an artery (red) and a vein (blue). \textbf{e}, The original (upper row) and FDD images (lower row) of the right femoral artery at different time points. \textbf{f, g}, SBR and SNR of both the original and FDD amplitude images. The signals are the averages of the maximum data from the two ears of each mouse. The backgrounds are the averages of ROIs selected from relatively flat areas outside the mice bodies. \textbf{h}, Amplitude of the origin of right femoral artery versus time. \textbf{i}, The derivative of \textbf{h}.}\label{fig6}
\end{figure}

\newpage
\setcounter{figure}{0} 
\renewcommand{\thefigure}{Extended Data Fig. \arabic{figure}} 
\renewcommand{\figurename}{} 

\begin{figure}[H]
\centering
\includegraphics[width=0.9\textwidth]{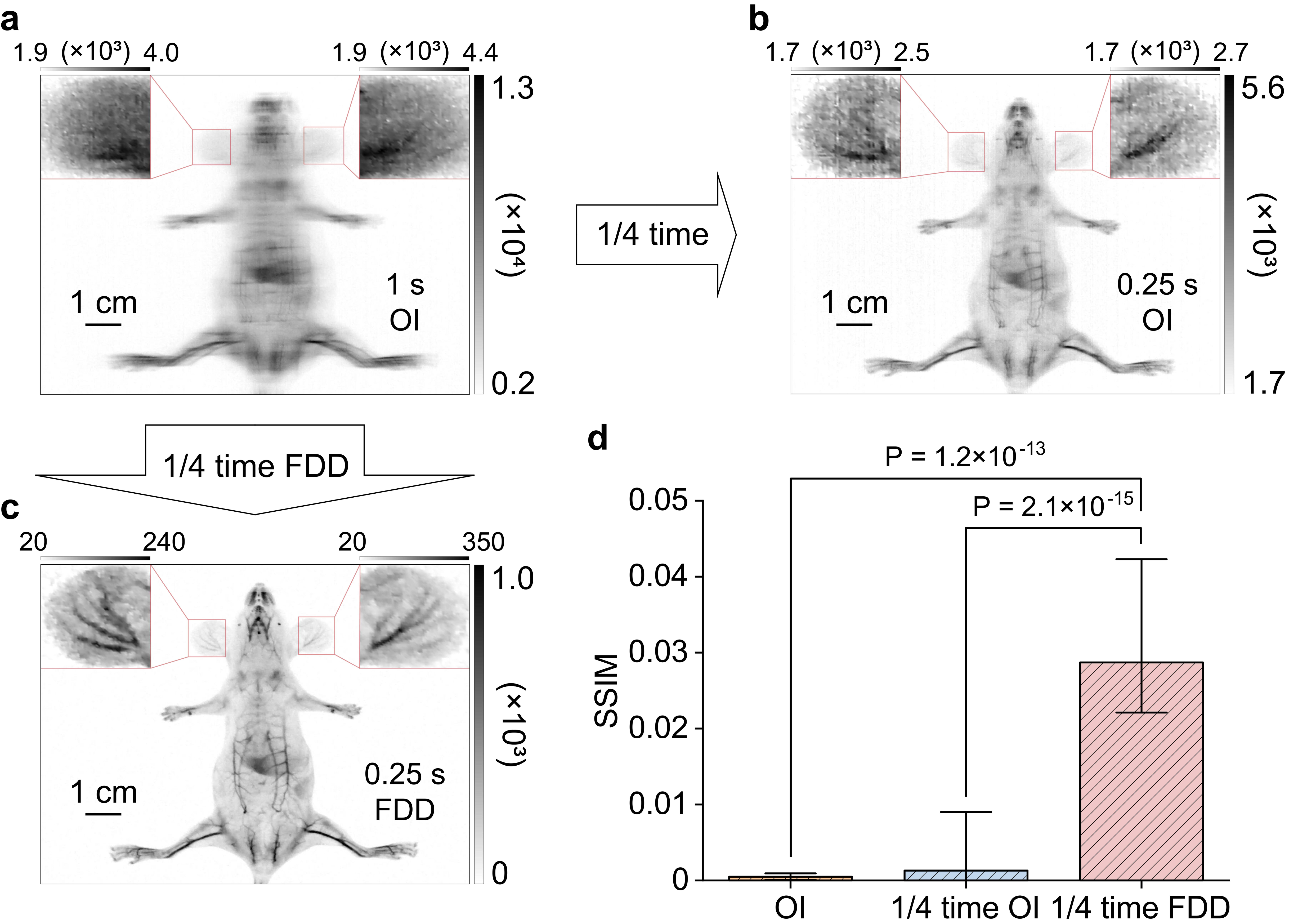} 
\caption{\textbf{Fluorescence imaging of whole-body vascular in a moving mouse under different exposure times. a}, The original image with severe motion blur. Exposure time = 1 s. \textbf{b}, The original image with reduced motion blur. Exposure time = 0.25 s. \textbf{c}, The FDD image without motion blur. Exposure time = 0.25 s. \textbf{d}, SSIM of original images and FDD images.}\label{fig7}
\end{figure}

\renewcommand{\thefigure}{\arabic{figure}} 

\end{document}